\documentstyle[12pt,twoside,fleqn,espcrc1]{article}
\newcommand{\be}{\begin{equation}}
\newcommand{\ee}{\end{equation}}
\newcommand{\ba}{\begin{eqnarray}}
\newcommand{\ea}{\end{eqnarray}}
\newcommand{\la}{\langle}
\newcommand{\ra}{\rangle}
\newcommand{\Mn}{ M_{\mbox{\tiny N}}}
\newcommand{\cGamma}{c_{\mbox{\tiny $\Gamma$}}}
\newcommand{\di}{ {\rm d} }
\newcommand{\bDelta}{ {{\mbox{\boldmath$\Delta$}} }}
\newcommand{\fslash}[1] {{\not\! #1\,}}

\newcommand{\singlesum}[1]{\!\!\!\!\sum_{\renewcommand{\arraystretch}{0.3}
	\begin{array}{l}\scriptscriptstyle{#1}\\
	\scriptscriptstyle {\phantom{n,\rm occ}}\end{array}} \!\!}
\newcommand{\doublesumUp}[3]{\!\!\!\!\sum_{\renewcommand{\arraystretch}{0.5}
	\begin{array}{c}\scriptscriptstyle {#1} \\ \scriptscriptstyle {#2} 
	\end{array}}^{\scriptscriptstyle{#3}} \!\!}
\newcommand{\triplelim}[3]
      {{\lim_{\renewcommand{\arraystretch}{0.5}
	\begin{array}{c}\scriptscriptstyle {#1} \\
        	        \scriptscriptstyle {#2} \\
			\scriptscriptstyle {#3}\end{array}} \!\!}}
\title{	\bf Polynomiality of off-forward distribution functions\\
	in the chiral quark soliton model\thanks{Presented 
	by P.~Schweitzer at the ``{ European Workshop on the QCD Structure 
	of the Nucleon}'' (QCD-N'02), Ferrara, Italy, 3-6 Apr 2002.}}
\author{P.~Schweitzer$^{\rm a}$\thanks{Supported by the contract 
	HPRN-CT-2000-00130 of the European Commission.},
	S.~Boffi$^{\rm a,b}$, M.~Radici$^{\rm a,b}$\\ 
 	\footnotesize\vspace{0.3cm}
 	$^{\rm a}$ Dipartimento di Fisica Nucleare e Teorica, 
	           Universit\`a di Pavia, I-27100 Pavia, Italy\\
 	$^{\rm b}$ Istituto Nazionale di Fisica Nucleare, 
	           Sezione di Pavia, I-27100 Pavia, Italy}
\begin{document}
\maketitle
\begin{abstract}
	\noindent
	Mellin moments of off-forward distribution functions are, 
	at $t=0$, even polynomials of the skewedness parameter $\xi$.
	It is proven that the unpolarized off-forward distribution functions 
	in the chiral quark soliton model satisfy this so called 
	polynomiality property. The proof is an important contribution to 
	the demonstration that the description of off-forward distribution 
	functions in the model is consistent.
\end{abstract}
\setcounter{footnote}{0}

\paragraph{Introduction}
Off-forward parton distribution functions (OFDFs) -- see 
\cite{Ji:1998pc,Radyushkin:2000uy,Goeke:2001tz} for recent reviews -- 
are a promising source of new information on the internal nucleon structure.
The understanding of non-perturbative properties of OFDFs 
-- which at present relies on models -- is essential in order to 
interpret first data on deeply virtual Compton scattering 
\cite{Saull:1999kt,Adloff:2001cn,Airapetian:2001yk,Stepanyan:2001sm,Delia},
or to make predictions for future experiments 
\cite{Kivel:2000fg,Korotkov:2001zn}.
Important contributions to our intuition on  non-perturbative aspects of 
OFDFs are based on calculations in the chiral quark soliton model ($\chi$QSM) 
\cite{Petrov:1998kf,Penttinen:1999th}.
The model has been derived from the instanton model of the QCD vacuum
\cite{Diakonov:2000pa}.
It describes -- without free adjustable parameters -- nucleon properties,
like form factors \cite{Christov:1995vm} and 
forward quark {\sl and} antiquark distribution functions 
\cite{Diakonov:1996sr,Diakonov:1997vc}, typically within (10-30)$\%$.

The reliability of the $\chi$QSM, however, is not only based on its
phenomenological success. More important -- from a theoretical point 
of view -- is the fact that it is possible to {\sl prove analytically}
that the model description of the nucleon is consistent. E.g., in 
ref.\cite{Diakonov:1996sr} it has been proven that forward quark 
and antiquark distribution functions in the model satisfy all general 
requirements such as sum rules, positivity and inequalities.
With the same rigour it has been shown in
refs.\cite{Petrov:1998kf,Penttinen:1999th}, that 
the model expressions for OFDFs reduce to usual parton distributions 
in the forward limit, and that their first moments yield form factors.
In this note a further contribution is made which demonstrates the
consistency of the $\chi$QSM. It is proven -- or rather the proof sketched 
-- that the model expression for $(H^u+H^d)(x,\xi,t)$ satisfies 
{\sl polynomiality}, i.e. the property that the $m^{\rm th}$ moment in $x$ 
of an OFDF at $t=0$ is an even polynomial in $\xi$ of degree less than or 
equal to $m$. In QCD, this property follows from Lorentz invariance.

\paragraph{OFDFs in the chiral quark soliton model}
The $\chi$QSM is based on an effective chiral low-energy field theory
with explicit quark, antiquark ($\bar\psi$, $\psi$) and Goldstone boson 
(for flavour SU(2) pion $\pi^a$) degrees of freedom. The effective theory 
is valid for energies below $600\,{\rm MeV}$ and given by the partition 
function
\be\label{eff-theory}
	Z_{\rm eff} 
	= \int\!\!{\cal D}\psi\,{\cal D}\bar{\psi}\,{\cal D}\pi\;
	\exp\Biggl(i\int\!\!\di^4x\;\bar{\psi}\,
	(i\fslash{\partial}-M\,U^{\gamma_5})\,\psi\Biggr)\;\;  , 
	\;\;\; U^{\gamma_5} = e^{i\gamma_5\tau^a\pi^a}\,. \ee
$M$ is the effective quark mass due to spontaneous breakdown of chiral 
symmetry.
In the limit of a large number of colours $N_c$ the nucleon emerges 
as a classical soliton of the pion field in the effective theory 
eq.(\ref{eff-theory}). In leading order of the large $N_c$ limit the
soliton field is static and one can determine the spectrum of the 
one-particle Hamiltonian 
\be\label{eff-Hamiltonian}
	\hat{H}_{\rm eff} \Phi_n({\bf x})  = E_n \Phi_n({\bf x}) \;,\;\;\;
	\hat{H}_{\rm eff} = 
	- i\gamma^0\gamma^k\partial_k + M\gamma^0U^{\gamma_5}\; \ee
of the effective theory eq.(\ref{eff-theory}).
The spectrum consists of a discrete bound state level and upper and lower 
Dirac continua. 
The nucleon is obtained by
occupying the bound state and the states of the lower continuum each by 
$N_c$ quarks in an anti-symmetric colour state, and considering the
zero modes of the soliton solution. 

The model allows to express nucleon matrix elements of QCD quark bilinear 
operators in terms of the single quark wave-functions $\Phi_n({\bf x})$.
In leading order of the large $N_c$ limit
\ba\label{matrix-elements}
	\la{\bf P'}|\bar{\psi}(z_1)\Gamma\psi(z_2)|{\bf P }\ra
	= \cGamma\,2\Mn N_c\singlesum{n, \rm occ}
	\int\!\!\di^3{\bf x}\:e^{i({\bf P'\!-P}){\bf x}}\,
	\bar{\Phi}_n({\bf z}_1\!-\!{\bf x})\Gamma\Phi_n({\bf z}_2\!-\!{\bf x})
	\,e^{iE_n(z^0_1-z^0_2)} \; ,\ea
where $\Gamma$ denotes some some Dirac- and flavour-matrix (which determines 
the constant $\cGamma$) and the sum goes over occupied states. 
$1/N_c$ corrections to eq.(\ref{matrix-elements}) can be taken systematically
into account. 
Eq.(\ref{matrix-elements}) enables one to evaluate in the model, e.g.
\ba\label{def-1}
&&	\mbox{\hspace{-0.7cm}}
	\int\!\frac{\di\lambda}{2\pi}\,e^{i\lambda x}\la {\bf P'},S_3'|
	\bar{\psi}_f(-\lambda n/2)\fslash{n}\psi_f(\lambda n/2)|{\bf P},S_3\ra
	\nonumber\\
&&	\mbox{\hspace{-0.8cm}}
	= H^f(x,\xi,t)\;\bar{U}({\bf P'},S_3')\fslash{n}U({\bf P},S_3)
 	+ E^f(x,\xi,t)\;\bar{U}({\bf P'},S_3')\,
	\frac{i\sigma^{\mu\nu}n_\mu\Delta_\nu\!}{2\Mn}\,U({\bf P},S_3) \;,\ea
which defines the twist-2 unpolarized OFDFs \cite{Ji:1998pc}. 
In eq.(\ref{def-1}) $n^\mu$ satisfies $n^2 = 0$ and $n(P'+P) = 2$.
The four-momentum transfer is defined as $\Delta^\mu = (P'-P)^\mu$, 
the skewedness parameter as $\xi = -\Delta n/2$, 
and the Mandelstam variable as $t = \Delta^2$.

In the large $N_c$ limit the nucleon mass $\Mn$ is ${\cal O}(N_c)$
but $|P^i|$, $|{P^i}'|$ are ${\cal O}(N_c^0)$, thus 
$t\!=\!-\bDelta^2 \!=\! {\cal O}(N_c^0)$.
The variables $x$, $\xi$ are both ${\cal O}(N_c^{-1})$.
With the convenient choice $n^\mu\! = \!(1,-{\bf e}^3)/\Mn$ one has
$\xi\! = \!-\Delta^3/2\Mn$. The spin-non-flip OFDF is given by
\ba\label{def-Hu+Hd-model}
	(H^u+H^d)(x,\xi,t)&=&\Mn N_c \int\!\!\di^3{\bf x}\;e^{i\bDelta{\bf x}}
	\sum\limits_{n,\,\rm occ} 
	\int\!\frac{\di z^0}{2\pi}\, e^{iz^0(x\Mn - E_n)}\nonumber\\
	&&\times\phantom{\biggl|}
	\Phi^{\!\ast}_n({\bf x}+{\textstyle\frac{z^0}{2}}{\bf e^3})\,
	(1+\gamma^0\gamma^3)\,
	\Phi_n({\bf x}-{\textstyle\frac{z^0}{2}}{\bf e^3})\ea
in LO of the large $N_c$ limit, while the flavour-nonsinglet combination 
$(H^u\!-\!H^d)(x,\xi,t)$ vanishes at this order.
Eq.(\ref{def-Hu+Hd-model}) has been derived and numerically evaluated in 
ref.\cite{Petrov:1998kf}.
The spin-flip OFDF $E^f(x,\xi,t)$ will be discussed elsewhere.

\paragraph{Sketch of the proof of polynomiality}
In this section the proof is sketched that the model expression for
$(H^u+H^d)(x,\xi,t)$, eq.(\ref{def-Hu+Hd-model}), satisfies polynomiality.
For the detailed proof see ref.\cite{polynomiality-paper}.
From eq.(\ref{def-Hu+Hd-model}) one obtains the model expression for the 
$m^{\rm th}$ moment $M^{(m)}_H(\xi,t)=\int\di x\,x^{m-1}(H^u+H^d)(x,\xi,t)$ 
for physical values of $t$.
In eq.(\ref{def-Hu+Hd-model}) the dependence of $M^{(m)}_H(\xi,t)$ on 
$\xi$ and $t$ is entirely determined by $\bDelta$ which appears in the
exponential $\exp(i\bDelta{\bf X})$ in eq.(\ref{def-Hu+Hd-model}).
The latter can be continued analytically to $t = -\bDelta^2 \to 0$ as
\be\label{Hu+Hd-an-cont}
	\triplelim{\rm analytical}{\rm continuation}{t\to 0}
	\exp(i\bDelta{\bf X}) =
	\sum\limits_{l_e=0}^\infty\,
	\frac{(-2i\xi\Mn|{\bf X}|\,)^{l_e} P_{l_e}}{l_e!\,} \ee
with $P_{l_e}$ denoting Legendre polynomials $P_{l_e}(\cos\theta)$.
This yields for the moments at $t=0$
\ba\label{Hu+Hd-model-mom2}
	M^{(m)}_H(\xi,0)
	&=&	\frac{N_c}{\Mn^{m-1}} \sum\limits_{n,\,\rm occ}
		\sum\limits_{k=0}^{m-1}\left(\matrix{ m-1\cr k}\right) 
		\frac{E_n^{m-1-k}\!\!\!}{2^k}\;
		\sum\limits_{j=0}^k\left(\matrix{k\cr j}\right)
		\doublesumUp{l_e=0}{l_e\,\rm even}{\infty}\,
		\frac{(-2i\xi\Mn)^{l_e}}{l_e!\,} \nonumber\\
	&&\times\la n|(\gamma^0\gamma^3)^k\,(\hat{p}^3)^{j}
		|\hat{\bf X}|^{l_e}\,P_{l_e}(\cos\hat{\theta})
		\,(\hat{p}^3)^{k-j}|n\ra \;.\ea
The ``ket'' states $|n\ra$ are related to 
coordinate space wave functions by $\Phi_n({\bf x})=\la{\bf x}|n\ra$, 
and $(\gamma^0\gamma^3)^k$ (equal to unity for even $k$ and to 
$\gamma^0\gamma^3$ for odd $k$) is introduced to simplify notation.
Finally, $\hat{p}^3$ is the free-momentum operator, which was 
``implicitly present'' in eq.(\ref{def-Hu+Hd-model}) due to
$\Phi_n({\bf x}-\frac{z^0}{2}{\bf e^3})=\la{\bf x}|e^{-iz^0\hat{p}^3}|n\ra$.
In $M^{(m)}_H(\xi,0)$, eq.(\ref{Hu+Hd-model-mom2}), only even powers
of $\xi$ appear: Odd powers of $\xi$ vanish due to symmetries of the model. 
What remains to be shown is that the series in $\xi^2$ terminates at
an $l_e^{\rm max}\le m$ for the $m^{\rm th}$ moment $M^{(m)}_H(\xi,0)$.

To see this note that the Hamiltonian $\hat{H}_{\rm eff}$ 
eq.(\ref{eff-Hamiltonian}) commutes with the grand-spin operator $\hat{\bf K}$ 
defined as sum of quark orbital angular momentum, spin and isospin.
In other words, simultaneous rotations in 3-space and isospin-space
-- around some axis ${\bf n}$ and an angle $\alpha$ 
   given by $\hat{R}=\exp(i\alpha\hat{\bf K})$ -- 
leave $\hat{H}_{\rm eff}$ invariant.
Under these rotations $\hat{H}_{\rm eff}$ and $|\hat{\bf X}|$ 
transform as rank 0,  $\gamma^0\gamma^3$ and $\hat{p}^3$ as rank 1, and
$\hat{P}_{l_e}(\cos\hat{\theta})\propto Y_{l_e 0}(\hat{\Omega})$ as rank $l_e$
irreducible tensor operators. 
An operator $\hat{T}^{L}_{M}$ is said to be an irreducible tensor operator
of rank $L$, if it transforms as $\hat{R}\,\hat{T}^{L}_M\,\hat{R}^\dag = 
\sum_{M'}D^{(L)}_{M'M}(R)\;\hat{T}^{L}_{M'}$ where $D^{(L)}_{M'M}(R)$ 
are finite rotations Wigner matrices.
The single quark states -- which are simultaneously eigenfunctions of 
$\hat{H}_{\rm eff}$, $\hat{\bf K}^2$ and $\hat{K}^3$, i.e. 
$|n\ra\equiv |E_n,K,M\ra$ -- transform as 
$\hat{R}|E_n,K,M\ra=\sum_{M'}D^{(K)}_{M'M}(R)\;|E_n,K,M'\ra$.

The product of two irreducible tensor operators $\hat{T}^{L'}_{M'}$ and 
$\hat{T}^{L''}_{M''}$ is a sum of some other irreducible tensor operators 
$\hat{T}^{L}_{M}$ with ranks $|L'\!-\!L''| \!\le\!  L \!\le\!  L'\!+\!L''$.
So the product of the irreducible tensor operators sandwiched 
between $\la n|...|n\ra$ in eq.(\ref{Hu+Hd-model-mom2}) is a sum of 
irreducible tensor operators with ranks $L$ ranging between 
$0 \le L \le k+l_e+1$ for odd $k$, and $0\le L\le k+l_e$ for even $k$.
Thus, in eq.(\ref{Hu+Hd-model-mom2}) one deals with traces 
(sums over matrix elements diagonal in $K$ and $M$)
of irreducible tensor operators.
The trace of an irreducible tensor operator, however, 
vanishes unless the operator has rank zero \cite{Fano-Racah}.

So $l_e$ in eq.(\ref{Hu+Hd-model-mom2}) cannot take arbitrary values, 
but is bound as
\be\label{Hu+Hd-Wig-Eck-3}
	l_e \le l_e^{\rm max}(k) = \cases{k+1 & for odd  $k$, \cr
		         		   k  & for even $k$.}\ee
If $l_e$ were larger than this $l_e^{\rm max}(k)$, it would be impossible
to compensate the rank $l_e$ of $P_{l_e}$ to obtain a rank zero operator,
even if the ranks of the 
other operators in eq.(\ref{Hu+Hd-model-mom2}) -- 
$(\gamma^0\gamma^3)^k$, $(\hat{p}^3)^j$, $(\hat{p}^3)^{m-j-1}$ --
would all add up.
Inserting the result eq.(\ref{Hu+Hd-Wig-Eck-3}) into 
eq.(\ref{Hu+Hd-model-mom2}) yields the desired result:
The $m^{\rm th}$ moment of $(H^u\!+\!H^d)(x,\xi,t)$ at $t\!=\!0$,
$M^{(m)}_H(\xi,0)$, is an even polynomial in $\xi$ of degree less than or
equal to $m$, i.e.
\be\label{def-polynom-Hu+Hd}
	\int\limits_{-1}^1\!\!\di x\:x^{m-1}\,(H^u+H^d)(x,\xi,0)
	= h^{(m)}_0 + h^{(m)}_2 \xi^2 + \dots + 
	    \cases{h^{(m)}_m    \xi^m     & for even $m$, \cr
		   h^{(m)}_{m-1}\xi^{m-1} & for odd  $m$} \ee
with coefficients $h^{(m)}_i$, $i=0,\,2,\,4,\,\dots\,\le m$,
explicitly given in eq.(\ref{Hu+Hd-model-mom2}).

\paragraph{Conclusions}
The proof has been sketched, that the chiral quark soliton model expression 
for the OFDF $(H^u+H^d)(x,\xi,t)$ satisfies polynomiality. 
The method can be generalized to prove polynomiality for other OFDFs 
in the model.
As a byproduct analytical expressions for moments at $t=0$ have been derived.
This opens the possibility, e.g., to evaluate the phenomenologically
particularly interesting coefficients in the Gegenbauer expansion of
the $D$-term \cite{Polyakov:1999gs,Belitsky:2000vk,Teryaev:2001qm}.
These coefficients have been extracted in \cite{Kivel:2000fg} from model 
results at physical values of $t$ by numerical extrapolation to $t=0$,
but with the results reported here they can be evaluated 
{\sl directly} at the unphysical point $t=0$.

\end{document}